\documentclass[twocolumn]{elsarticle}    

\usepackage{amsmath}
\usepackage{amssymb}
\usepackage{graphicx}

\begin{document}

\begin{frontmatter}

\title{Discrete-Time Nonlinear Systems Identification with Probabilistic Safety and Stability Constraints\thanksref{footnoteinfo}} 

\thanks[footnoteinfo]{This work was supported in part by NSF grant no. SMA-2134367 and in part by a Space Technology Research Institutes grant (number 80NSSC19K1076) from NASA Space Technology Research Grants Program.}

\author[a]{Iman Salehi}\ead{iman.salehi@uconn.edu},    
\author[a]{Tyler Taplin}\ead{tyler.taplin@uconn.edu},               
\author[a]{Ashwin Dani}\ead{ashwin.dani@uconn.edu}  

\address[a]{Department of Electrical and Computer Engineering, University of Connecticut, Storrs, CT, USA}  


\begin{abstract}                          
This paper presents a discrete-time nonlinear system identification method while satisfying the stability and safety properties of the system with high probability. An Extreme Learning Machine (ELM) is used with a Gaussian assumption on the function reconstruction error. A quadratically constrained quadratic program (QCQP) is developed with probabilistic safety and stability constraints that are only required to be satisfied at sampled points inside the invariant region. The proposed method is validated using two simulation examples: a two degrees-of-freedom (DoF) robot manipulator with constraints on joint angles whose trajectories are guaranteed to remain inside a safe set and on motion trajectories data of a hand-drawn shape.
\end{abstract}

\end{frontmatter}

\section{Introduction}
Data-driven dynamic system model learning approaches create black box models that do not require much prior first principles knowledge about the system \cite{brunton2016discovering,chiuso2019system,sznaier2020control}. 
The data generated by the controlled system often exhibit system properties such as stability, invariance with respect to a set and so on. Preserving these properties in system identification by formulating them as constraints leads to accurate identification of dynamical systems \cite{chiuso2019system,sznaier2020control}. This can be useful in many model learning applications to engineering problems \cite{maciejowski1995guaranteed,khosravi2020nonlinear,khansari2014learning,ravichandar2017learning,salehi2019active,umlauft2020learning}. In \cite{ravichandar2017learning,khansari2014learning} and \cite{singh2021learning} stability and stabilizability of system dynamics is considered for model learning in learning from demonstration. Our recent efforts have focused on adding safety property in the data-driven system model learning, see, e.g., \cite{salehi2019active,salehi2021dynamical}.

The concept of safety is centered around the idea of constraining the behavior of the system model to a prescribed set, by ensuring forward invariance of the set with respect to the system model. Barrier functions (BF) are commonly used to certify the forward invariance of a closed set with respect to a system model. 

Control barrier functions (CBFs) are used for the synthesis of safety-critical controllers via Quadratic Programming (QP) \cite{nguyen2016exponential,berkenkamp2016safe,ames2017control}. Control Lyapunov functions (CLF) and CBF are merged to synthesize stable and safe controllers by solving a QP in \cite{romdlony2016stabilization,xu2015robustness}. In \cite{taylor2019adaptive}, an adaptive CBF (aCBF) is proposed, which ensures the forward invariance of a closed set with respect to a nonlinear control-affine system with parametric uncertainties. 
In \cite{Lopez2021robust} the aCBF method is merged with an adaptive data-driven safety controller for contracting systems. These methods mainly focus on controller synthesis problem with a known dynamic system model. The controller synthesis methods assume the knowledge of barrier function. A framework for estimating barrier functions and safe regions directly from sensor data is developed in \cite{berkenkamp2016safe} and  \cite{srinivasan2020synthesis}.

Classical system identification techniques, focused on linear systems, utilize parametric model structure and guarantee the stability of learned models \cite{chiuso2019system,sznaier2020control}. Machine learning techniques, such as neural networks and Gaussian processes (GPs), have recently been used for nonlinear system identification. 
Model uncertainties, in the form of the neural network approximation error and external disturbances, are accounted for in our prior study in \cite{salehi2021dynamical} along with stability and safety constraints using reciprocal BFs, but noise is not explicitly modeled. In the context of Bayesian learning, GP state-space models are learned with stability constraints in \cite{umlauft2020learning}. In \cite{khojasteh2020probabilistic}, simultaneous continuous system model learning and control synthesis method is developed, where GPs are used for model learning, and probabilistic CBF and CLF are used to achieve safety and stability. This paper develops a system identification method for discrete-time systems while simultaneously verifying probabilistic safety and stability properties by utilizing CBFs for discrete-time nonlinear systems developed in \cite{agrawal2017discrete}. The uncertainties are modeled using white Gaussian noise. A discrete-time zeroing control barrier function (DT-ZCBF) and discrete-time control Lyapunov functions (DT-CLF) are used to obtain safety and stability chance constraints.
An extreme learning machine (ELM) network, which is a computationally efficient approximation for system model, is used to represent the unknown nonlinear discrete-time system. The ELM parameter learning problem is formulated as a quadratically constrained quadratic program (QCQP) subject to discrete-time safety and stability chance constraints. The optimization problem contains finite number of decision variables with infinite constraints. To make the optimization tractable, an equivalent optimization problem is derived that requires the constraints to be evaluated only at a finite number of sampled points. Finally, the constrained model learning approach is demonstrated in simulation using two examples: a $2$ DoF planar robot whose joint positions are lower and upper bounded using an ellipse zeroing barrier function, and learning motion trajectories of one of the complicated shapes from a publicly available dataset \cite{lemme2015open}. 

\section{Preliminaries}\label{sec:prelim}
In this section, preliminaries of discrete-barrier and discrete Lyapunov functions are presented. Consider an unknown nonlinear discrete-time system of the form
\begin{equation}
     x(k+1)=f(x(k),u(k)),\label{eq:orig_dyn}
\end{equation}
where $f:\!\mathcal{D}\times\mathcal{U} \rightarrow \mathcal{D}$ is a continuous function, $x(k) \in \mathcal{D} \subset \mathbb{R}^n$ is the state of the system at time step $k\in \mathbb{Z}^+$, and $u(k) = \pi(e(k))\in \mathcal{U} \subset \mathbb{R}^{m}$ is the system control input with a policy $\pi(\cdot)$ that depends on the error $e(k) = x(k)-x^*$, and $x^*$ is the system equilibrium. For ease of notation, the time dependency of variables, for example, $x(k)$ is abbreviated by $x_k$, unless necessary for clarity.

\subsection{Discrete-Time Control Lyapunov Functions}\label{subsec:DLF}
The classical stability analysis of continuous-time nonlinear dynamical systems using Lyapunov theory is extended to the discrete-time domain. It entails finding a positive definite or semi-definite function $V:\!\ \mathcal{D}\rightarrow \mathbb{R}$, which decreases along the trajectories of the system given in (\ref{eq:orig_dyn}). Refer to \cite{agrawal2017discrete} and references therein for a comprehensive study on discrete-time Lyapunov theorems. 
\begin{defn}[DT-CLF] \label{def:1}\normalfont
Given the discrete-time system (\ref{eq:orig_dyn}), $V(x_k)$ is said to be a DT-CLF if it can be bounded by $\alpha_1(\Vert x_k \Vert) \leq V(x_k) \leq \alpha_2(\Vert x_k\Vert)$ and there exists a control input $u_k\!:\mathcal{D}\rightarrow\mathcal{U}$ such that
\begin{equation}
    \mathcal{C}_L\left(x_k,u_k\right)=\Delta V(x_{k},u_k) + \beta(\Vert x_k\Vert) \leq 0, \ \forall x_k\in \mathcal{D}, \label{eq:DLF}
\end{equation}
where $\Delta V(x_{k},u_k):=V(x_{k+1})-V(x_{k})=\Delta V_k$, $\alpha_1,\ \alpha_2,$ and  $\beta$ belong to class $\mathcal{K}$ function, and $\Vert \cdot \Vert$ is an arbitrary norm on $\mathbb{R}^n$.
\end{defn}

\subsection{Discrete-Time Zeroing Control Barrier Functions}\label{subsec:DBF}

Safety properties can be achieved by constraining  the solutions of the discrete-time dynamical system to a  pre-specified closed set $\mathcal{S}\!=\!\{x_k\in\mathcal{D}:\!\ h(x_k)\geq0\}$, where $h\! :\mathbb{R}^n\rightarrow\mathbb{R}$ is a continuously differentiable function associated with a barrier function \cite{ames2017control}.

\begin{defn}[DT-ZCBF]\label{def:2}\normalfont
Given the discrete-time system (\ref{eq:orig_dyn}), the set $\mathcal{S}$, a continuously differentiable function $h\!:\mathbb{R}^n \rightarrow \mathbb{R}$ is said to be a DT-CZBF, if there exists a class $\mathcal{K}$ function satisfying $\eta(r) < r$, $\forall r>0$, a set $\mathcal{D}$ with $\mathcal{S} \subseteq \mathcal{D} \subset \mathbb{R}^n$, and a control input $u_k\!:\mathcal{D}\rightarrow\mathcal{U}$ such that
\begin{equation}
  \mathcal{C}_{\mathcal{B}}\left(x_k,u_k\right)\!=\!\Delta h(x_{k},u_{k})+ \eta(h(x_k))\!\geq\! 0,\  \forall x_k\!\in \mathcal{D},  \label{eq:DZBF-cond}
\end{equation} where $\Delta h(x_{k},u_{k}):=h(x_{k+1})-h(x_k) = \Delta h_k$.
\end{defn}

The set $\mathcal{S}$ is \textit{forward invariant} with respect to the discrete-time system in (\ref{eq:orig_dyn}) if and only if there exists a DT-ZCBF as defined in Definition \ref{def:2}, which means for any initial condition $x_0\in \mathcal{S}$, implies $x_k \in \mathcal{S}$, $\forall  k\in \mathbb{Z}^+$ \cite{ahmadi2019safe}.

\begin{assum}\label{assump_1} \normalfont
The function $f\!:\mathcal{D}\times \mathcal{U}\rightarrow\mathcal{D}$ is locally Lipschitz continuous and bounded in $\mathcal{D}$. 
\end{assum}

\section{Constrained ELM Learning Problem}\label{sec: Problem-Formulation}

Consider a discrete-time nonlinear dynamical system in (\ref{eq:orig_dyn}), it is assumed that the unknown function $f(\cdot)$ can be parametrized as 
$f\left(x_k,\pi(e_k)\right) = \textbf{W}^{T}g(s_k)$,
where $\textbf{W}\in\mathbb{R}^{\left(n_{h}+1\right)\times n}$ is the unknown ideal bounded constant output layer weight matrix, $g(s_k)\!=\![\psi(q\cdot s_k),1]\!:\mathbb{R}^{n+m+1}\!\rightarrow \mathbb{R}^{n_{h}+1}$,
$s_k\!=\!\left[x_k^{T},e_k^T,1\right]^{T}\in\mathbb{R}^{n+m+1}$,
$\psi(\cdot)\in \mathbb{R}^{n_h}$ is the vector Sigmoid activation function, $n_{h}$ is the number of neurons in the hidden layer of the ELM. The vector Sigmoid activation function $\psi(\cdot)$ is given by
\begin{align}
    \psi\left(q\cdot s_k\right)\!=\!\left[\frac{1}{1\!+\!e^{-\left(q\cdot s_k\right)_{1}}},\cdots,\frac{1}{1\!+\!e^{-\left(q\cdot s_k\right)_{n_{h}}}}\right]^{T},
 \label{eq:sigmoid}
\end{align}where $q=[P,\ b_{p}]\in\mathbb{R}^{n_{h}\times (n+m+1)}$,  $P\!=\!\mathrm{diag}(a_{p})\cdot U^{T}\in\mathbb{R}^{n_{h}\times (n+m)}$, $a_p\in\mathbb{R}^{n_h}$ and  $b_p\in \mathbb{R}^{n_h}$ are the internal slopes and biases vectors, respectively,  $U\in\mathbb{R}^{(n+m)\times n_{h}}$ is the ideal bounded input layer weight matrix. The parameters $a_p$, $b_p$ and $U$ of ELM are computed using intrinsic plasticity (IP) or batch intrinsic plasticity (BIP) algorithms. See \cite{salehi2021dynamical} for preliminaries of ELM. Other activation functions such as radial basis functions, tangent Sigmoid can also be used \cite{lewis2002neuro,huang2015trends}. 
Using the ELM parametrization of $f(\cdot)$, the discrete system in (\ref{eq:orig_dyn}) is written as 
\begin{align}
x_{k+1} & =W^{T}g(s_k)+\epsilon_k,\label{eq:compact_NN}
\end{align}where $W \in \mathbb{R}^{\left(n_h+1\right)\times n}$ is the estimated ELM output weight matrix and $\epsilon_k \in \mathbb{R}^n$ is the uncertainty due to unmodeled effects.

\begin{assum}\label{assump_2} \normalfont
The weights of the ELM are bounded by known positive constants, i.e., $\Vert W \Vert_{F} \leq \bar{W}$, $\Vert U \Vert_{F} \leq \bar{U}$, where $\Vert \cdot \Vert_F$ is the Frobenius norm \cite{lewis2002neuro}.
\end{assum}

\begin{rem}\label{rem:1}\normalfont
The sigmoid function $\psi_i(\cdot)\in [0,1]$ and hence its derivative $\psi_i(\cdot)(1-\psi_i(\cdot))$ has upper and lower bounds given by $0\leq\psi_i(\cdot)(1-\psi_i(\cdot))\leq0.25$, $\forall i = 1,\cdots n_h$. 
Thus, $\Vert \psi(\cdot) \Vert \leq \sqrt{n_h}$ and $\Vert \psi(\cdot)(1-\psi(\cdot))\Vert 
\leq 0.25\sqrt{n_h}$. Using Assumptions \ref{assump_2}, Assumption  \ref{assump_1} is still valid for the ELM parameterization of $f(x_k,u_k)$ \cite{lewis2002neuro}.
\end{rem}

\begin{assum}\label{assump_3}\normalfont
The uncertainty, $\epsilon_{k}$ is independent and identically distributed with Gaussian distribution of zero mean and covariance $\sigma^2I$, i.e., $\epsilon_{k} \sim \mathcal{N}\left(0,\sigma^2I\right)$ \cite{muller1998issues,Rasmussen2004}. \end{assum}

Let $\textsc{X}_{1:T_m}\!\!=\!\!\{x_1,\cdots,x_{T_m}\}$ denote $T_m$ time samples of the system's trajectory of (\ref{eq:orig_dyn}) and $\textsc{G}_{0:T_{m}-1}=\{g(s_0),\cdots,g(s_{T_m-1})\}$ denote their Sigmoid functions.
The Gaussian assumption on the ELM reconstruction error for the dynamic model in (\ref{eq:compact_NN}) yields the following likelihood function
\begin{align}
p\!\left(\textsc{X}\vert \textsc{G},W\right)\! & =\prod_{k=0}^{T_{m}-1}\!\!\!\frac{1}{(2\pi)^{\frac{n}{2}} \sigma^n}\mathrm{exp}\!\!\left(\!-\!\frac{\Vert x_{k+1}-W^{T}g(s_k)\Vert^{2}}{2\sigma^{2}}\!\right),\nonumber \\
& = \prod_{k=0}^{T_m-1} \mathcal{N}(W^Tg(s_k),\sigma^2I)\label{eq:likelihood}
\end{align}
where the indexes for $\textsc{X}$ and $\textsc{G}$ are dropped for brevity. The Gaussian distribution of the ELM reconstruction error also induces a distribution over the DT-CLF and DT-ZCBF constraints in (\ref{eq:DLF}) and (\ref{eq:DZBF-cond}), which can be used to obtain probabilistic stability and safety guarantees in the constrained parameter learning of the ELM, respectively. \\
Given a set of state trajectories and state feedback control, the constrained system identification problem is formulated as a parameter learning problem of an ELM approximation of  $f(\cdot)$ given in (\ref{eq:compact_NN}), such that the system in (\ref{eq:compact_NN}) satisfy probabilistic safety and stability properties.

\section{Chance-constrained Optimization}
\subsection{Encoding Probabilistic Safety Constraints}

DT-ZCBF defines a forward invariant set such that solutions of the nonlinear dynamical system that start in that set remain in that set for all time. A hyperellipsoid is used to represent a DT-ZCBF, $h(x_k)\!:\!\mathbb{R}^n \rightarrow \mathbb{R}$, and it is given by
\begin{align}
    h(x_k)=1-(x_k-\bar{x})^{T}A(x_k-\bar{x}), \label{eq:h_k}
\end{align}
where $A\in \mathbb{R}^{n\times n}$
is a symmetric positive definite matrix, and $\bar{x}$ is the vector that denotes the center of gravity of the hyperellipsoid.

\begin{lem}\label{lem:1}
Given the system model in (\ref{eq:compact_NN}) and Assumption \ref{assump_3}, the uncertainty, $\epsilon_k$ induces a distribution over the DT-ZCBF constraint in (\ref{eq:DZBF-cond}) with the following parameters:
\begin{align}
\mathbb{E}\left[\mathcal{C}_{\mathcal{B}}\right]\! &= \!\! 1\!-\!\left\Vert \!A^{\frac{1}{2}}\!\left(W^{T}\!g\!-\!\bar{x}\right)\right\Vert ^{2}\!\!\!\!\!-\!\sigma^2\mathrm{tr}\!\left(A\right)\!\!-\!(1\!-\!\gamma)h(x_{k})\label{eq:E_DZBF}\\
V\!ar\!\left[\mathcal{C}_{\mathcal{B}}\right] &=  4 \left\Vert \sigma A\left(W^{T}\!g\!-\!\bar{x}\right)\right\Vert ^{2}\!+\!2\sigma^4\mathrm{tr}\left(A^2\right)\!,&\label{eq:Var_DZBF}
\end{align}
where the class $\mathcal{K}$ function $\eta(h(x_k))$ in Definition \ref{def:2} is selected as $\eta(h(x_k))=\gamma h(x_k)$, and $\gamma$ is a positive constant.
\end{lem}
\begin{pf}
Substituting (\ref{eq:h_k}) in (\ref{eq:DZBF-cond}) for $h(x_{k+1})$ and using (\ref{eq:compact_NN}) yields 
\begin{align}
    \mathcal{C}_{\mathcal{B}}\!=\!\!1\!\!-\!\left(W^{T}\!g\!+\!\epsilon_{k}\!-\!\bar{x}\right)^{T}\!\!\!\!A\!\left(W^{T}\!\!g\!+\!\epsilon_{k}\!\!-\!\bar{x}\right)\!\!-\!\!\left(1\!\!-\!\!\gamma\right)\!h(x_{k}).
\end{align} Taking the expectation of $\mathcal{C}_{\mathcal{B}}$ we have
\begin{align}
    &\mathbb{E}\!\left[1\!-\!\left(W^{T}\!g\!+\!\epsilon_{k}\!-\!\bar{x}\right)^{T}\!\!\!A\!\left(W^{T}\!g\!+\!\epsilon_{k}\!-\!\bar{x}\right)\!-\!\left(1\!-\!\gamma\right)h(x_{k})\right] \!= \nonumber\\
    &1\!\!-\!\!\left(W^{T}\!\!g\!-\!\bar{x}\right)^{T}\!\!\!A\!\left(W^{T}\!\!g\!-\!\bar{x}\right)\!\!-\!\!\mathrm{tr}\!\left(A\sigma^2\right)\!\!-\!\!(1\!\!-\!\gamma)h(x_k),
\end{align} 
and the variance is
\begin{align}
V\!ar\!\left[\mathcal{C}_{\mathcal{B}}\right]\!=&\mathbb{E}\!\left[\!\left(h(x_{k+1})\!-\!\mathbb{E}\left[h(x_{k+1})\right]\right)^{2}\!\right]  \label{eq:lem1_var1}\\
\!\!=&\mathbb{E}\!\left[\!\left(-2g^{T}WA\epsilon_k\!-\!\epsilon_k^{T}A\epsilon_k+2\bar{x}^{T}A\epsilon_k\!+\!\mathrm{tr}(A\sigma^2)\right)\!^{2}\right]\!. \nonumber
\end{align}
Utilizing properties such as the odd moments of multivariate Gaussian distribution are zero and the trace when the matrices are of suitable dimensions is invariant under circular permutations, (\ref{eq:lem1_var1}) can be written as 
{blue}{\begin{align}
&V\!ar\left[\mathcal{C}_{\mathcal{B}}\right]=4\sigma^{2}\left(g^{T}WA^{2}W^{T}g+\bar{x}^{T}A^{2}\bar{x}-2g^{T}WA^{2}\bar{x}\right)+\nonumber\\
&\qquad \qquad \quad \mathbb{E}\left[\left(\epsilon_k^{T}A\epsilon_k\right)^{2}\right]-\sigma^{4}\mathrm{tr}(A)^{2} \label{eq:lem1_var2}
\end{align}}After substituting $\mathbb{E}\left[\left(\epsilon_k^{T}A\epsilon_k\right)^{2}\right]=\sigma^4 \mathrm{tr}(A)^2+2\sigma^4\mathrm{tr}(A^2)$ in (\ref{eq:lem1_var2}), the result in (\ref{eq:Var_DZBF}) is obtained.
\end{pf}

To account for the uncertainty in safety constraint satisfaction, the DT-ZCBF is rewritten as 
\begin{align}
    \mathbb{P}\left(\mathcal{C}_{\mathcal{B}}\geq \zeta \vert x_k\right)= 1 - \mathcal{F}_{\mathcal{C}_{\mathcal{B}}}\left(\zeta\right) \geq p_k \label{eq:P_CB}
\end{align}
where $p_k\in (0,1)$ is a user-specified risk tolerance, $\zeta \in \mathbb{R}^+$, and $\mathcal{F}_{\mathcal{C}_{\mathcal{B}}}(\zeta)$ is the cumulative distribution of the standard Gaussian given by
\begin{align}
    \mathcal{F}_{\mathcal{C}_{\mathcal{B}}}(\zeta)\!\triangleq\! \Phi\!\left(\!\frac{\zeta-\mathbb{E}\left[\mathcal{C}_{\mathcal{B}}\right]}{\sqrt{V\!ar\!\left[\mathcal{C}_{\mathcal{B}}\right]}}\!\right)\!\!=\! \frac{1}{2}\!+\!\frac{1}{2}\mathrm{erf}\!\left(\!\frac{\zeta-\mathbb{E}\left[\mathcal{C}_{\mathcal{B}}\right]}{\sqrt{2V\!ar\!\left[\mathcal{C}_{\mathcal{B}}\right]}}\!\right) \label{eq:erf_BF}
\end{align}
and  $\Phi(\cdot)$ is a cumulative distribution function of the standard Gaussian. The parameters derived in Lemma \ref{lem:1} can be used to compute cumulative distribution  in (\ref{eq:erf_BF}). Note that as the variance tends to zero, i.e., $V\!ar\left[\mathcal{C}_{\mathcal{B}}\right]\rightarrow 0 $, the probability $\mathbb{P}\left(\mathcal{C}_{\mathcal{B}}\geq \zeta \vert x_k\right)$ tends to one when $\zeta < \mathbb{E}\left[\mathcal{C}_{\mathcal{B}}\right]$. As the uncertainty in the system dynamics decreases to zero the safety assurance can be guaranteed with probability one. 

Substituting (\ref{eq:erf_BF}) into (\ref{eq:P_CB}), the probabilistic safety constraint is given by 
\begin{align}
    \zeta - \mathbb{E}\left[\mathcal{C}_{\mathcal{B}}\right] \leq -c(p_k)\sqrt{V\!ar\left[\mathcal{C}_{\mathcal{B}}\right]}, \label{eq:Prob-DTZBF}
\end{align}
where $c(p_k)=\sqrt{2}\ \mathrm{erf}^{-1}\left(2p_k-1\right)$.

\subsection{Encoding Probabilistic Stability Constraints}

In this subsection, a probabilistic stability constraint is derived using Definition \ref{def:1} for the uncertain discrete-time system given in (\ref{eq:compact_NN}) and a Lyapunov candidate $V(x_k)=(x_k-x^*)^TP(x_k-x^*)$, where $P\in \mathbb{R}^{n\times n}$ such that $P=P^T>0$.
\begin{lem}\label{lem:2}
Given the system in (\ref{eq:compact_NN}) and Assumption \ref{assump_3}, the Gaussian model uncertainty induces a distribution over the DT-CLF constraint in (\ref{eq:DLF}) with the following parameters:
\begin{align}
\mathbb{E}\left[\mathcal{C}_{L}\right]\!= & \left\Vert P^{\frac{1}{2}}\!\!\left(W^{T}\!g\!-\!x\!^*\!\right)\right\Vert ^{2}\!\!\!\!+\!\sigma^2\mathrm{tr}\!\left(P\right)\!-\!(1\!-\!\rho)V(x_{k}),\!\!\label{eq:E_DLF}\\
V\!ar\!\left[\mathcal{C}_{L}\right]= & \ 4\left\Vert \sigma P\left(W^{T}\!g\!-\!x^*\!\right)\right\Vert ^{2}\!+\!2\sigma^4\mathrm{tr}\left(P^2\right),\label{eq:Var_DLF}
\end{align}
where the class $\mathcal{K}$ function $\beta(\Vert x_k \Vert)$ in Definition \ref{def:1} is defined as $\beta(\Vert x_k \Vert) = \rho V(x_k)$ and $\rho$ is a positive constant. \\
\end{lem}

\begin{pf}
Substituting $V(x_k)=(x_k-x^*)^TP(x_k-x^*)$ in (\ref{eq:DLF}) for $V(x_{k+1})$ and using (\ref{eq:compact_NN}) yields
\begin{align}
    \mathcal{C}_L\!\!=\!\!\left(W^{T}\!g\!+\!\epsilon_{k}\!-\!x^*\right)\!^{T}\!P\!\left(W^{T}\!\!g\!+\!\epsilon_{k}\!\!-\!x^*\right)\!\!-\!\!\left(1\!\!-\!\!\rho\right)\!V(x_{k}).
\end{align} Taking the expectation of $\mathcal{C}_L$ we have
\begin{align}
    &\mathbb{E}\!\left[\left(W^{T}\!g\!+\!\epsilon_{k}\!-\!x^*\right)^{T}\!\!\!P\!\left(W^{T}\!g\!+\!\epsilon_{k}\!-\!x^*\right)\!-\!\left(1\!-\!\rho\right)V(x_{k})\right] \!= \nonumber\\
    &\left(W^{T}\!\!g\!-\!x^*\right)^{T}\!\!\!P\!\left(W^{T}\!\!g\!-\!x^*\right)\!\!+\!\!\mathrm{tr}\!\left(P\sigma^2\right)\!\!-\!\!(1\!\!-\!\rho)V(x_k),
\end{align} 
and the variance is
\begin{align}
V\!ar\!\left[\mathcal{C}_L\right]\!=&\mathbb{E}\!\left[\!\left(V(x_{k+1})\!-\!\mathbb{E}\left[V(x_{k+1})\right]\right)^{2}\!\right]  \label{eq:lem2_var1}\\
\!\!=&\mathbb{E}\!\left[\!\left(2g^{T}WP\epsilon_k\!+\!\epsilon_k^{T}P\epsilon_k\!-\!2x^{*^T}\!P\epsilon_k\!-\!\mathrm{tr}(P\sigma^2)\right)^{2}\right]\!. \nonumber
\end{align}
Utilizing properties such as the odd moments of multivariate Gaussian distribution are zero and the trace when the matrices are of suitable dimensions is invariant under circular permutations, (\ref{eq:lem2_var1}) can be written as 
\begin{align}
&V\!ar\!\left[\mathcal{C}_L\right]=4\sigma^{2}\left(g^{T}WP^{2}W^{T}g\!+\!x^{*^{T}}\!\!P^{2}x^*\!-\!2g^{T}WP^{2}x^*\right)\!+\nonumber\\
&\qquad \qquad \quad \mathbb{E}\left[\left(\epsilon_k^{T}P\epsilon_k\right)^{2}\right]-\sigma^{4}\mathrm{tr}(P)^{2} \label{eq:lem2_var2}
\end{align}After substituting $\mathbb{E}\left[\left(\epsilon_k^{T}P\epsilon_k\right)^{2}\right]=\sigma^4 \mathrm{tr}(P)^2+2\sigma^4\mathrm{tr}(P^2)$ in (\ref{eq:lem2_var2}), the result in (\ref{eq:Var_DLF}) is obtained.
\end{pf}

To account for the uncertainty in stability constraint satisfaction, the DT-CLF in (\ref{eq:DLF}) can be rewritten as 
\begin{align}
    \mathbb{P}\left(\mathcal{C}_L\leq\delta \vert x_k\right) = \mathcal{F}_{\mathcal{C}_L}(\delta) \geq p_k, \label{eq:P_CL}
\end{align}
where 
\begin{align}
    \mathcal{F}_{\mathcal{C}_L}(\delta)=\frac{1}{2}+\frac{1}{2}\mathrm{erf}\left(\frac{\delta-\mathbb{E}\left[\mathcal{C}_{L}\right]}{\sqrt{2V\!ar\!\left[\mathcal{C}_{L}\right]}}\right). \label{eq:erf_LF}
\end{align}
Substituting (\ref{eq:erf_LF}) into (\ref{eq:P_CL}) and utilizing (\ref{eq:E_DLF}) and (\ref{eq:Var_DLF}), the probabilistic stability constraint is given by
\begin{align}
    \delta - \mathbb{E}\left[\mathcal{C}_L\right]\geq c(p_k)\sqrt{V\!ar\left[\mathcal{C}_L\right]},\label{eq:Prob-DTLF}
\end{align}
where $\mathbb{E}\left[\mathcal{C}_L\right]$ and $V\!ar\left[\mathcal{C}_L\right]$ are given in Lemma \ref{lem:2}.
\newline
\begin{assum}\label{assump_4}\normalfont
The variances $V\!ar\left[\mathcal{C}_{\mathcal{B}}\right]\geq 1$ and $V\!ar\left[\mathcal{C}_L\right]\geq 1$.
\end{assum}
\begin{rem}\label{rem:2}\normalfont
Functions $\sqrt{V\!ar\left[\mathcal{C}_{\mathcal{B}}\right]}$ and $\sqrt{V\!ar\left[\mathcal{C}_L\right]}$ are not Lipschitz continuous on $\left[0,1\right]$ as the slope of the tangent line to their corresponding arguments become steeper as they approach to zero. Therefore, Assumption \ref{assump_4} is made to satisfy the constraint's Lipschitz continuity requirement.
\end{rem}
\begin{rem}\label{rem:3}\normalfont
If Assumption \ref{assump_4} does not hold, i.e., $V\!ar\left[\mathcal{C}_{\mathcal{B}}\right] < 1$ and $V\!ar\left[\mathcal{C}_L\right] < 1$, the right hand side (RHS) of the constraints in (\ref{eq:Prob-DTZBF}) and (\ref{eq:Prob-DTLF}) can be modified by introducing a constant $\xi > 1$ such that $\sqrt{\xi^2 V\!ar[\mathcal{C}_{\mathcal{B}}]}\geq 1$ and $\sqrt{\xi^2 V\!ar[\mathcal{C}_L]}\geq1$. For example, for the safety constraint in (\ref{eq:Prob-DTZBF}) we get 
\begin{align}
    \zeta - \mathbb{E}\left[\mathcal{C}_{\mathcal{B}}\right]\leq \frac{-c(p_k)}{\xi}\sqrt{\xi^2V\!ar\left[\mathcal{C}_{\mathcal{B}}\right]},\label{eq:proposition_1}
\end{align}
which the right hand side of (\ref{eq:proposition_1}) can be lower bounded by $\zeta-\mathbb{E}\left[\mathcal{C}_{\mathcal{B}}\right]\leq-\xi c(p_{k})V\!ar\left[\mathcal{C}_{\mathcal{B}}\right]$. Scaling by the selected $\xi$ is equivalent to solving the optimization problem with a tighter constraint. \end{rem}

\subsection{Chance-Constrained ELM Learning Problem}

Given a safe set $\mathcal{S}$ defined in Section \ref{sec:prelim}, the ELM parameterized model in (\ref{eq:compact_NN}), the initial state $x_{0} \in \mathcal{S}$, and the likelihood function $p\!\left(\textsc{X}\vert \textsc{G},W\right)$ in (\ref{eq:likelihood}), the ELM parameter learning problem can be formulated as a chance constrained regularized maximum log-likelihood estimation as follows 
\begin{subequations}
\begin{align}
    W^* \!\!& =\arg\!\!\!\!\!\!\!\!\!\min_{W \in \mathbb{R}^{(n_{h}\!+\!1)\times n}}\!\!\!\!\!\!\!\!\! -\ln \ p\left(\textsc{X}\vert \textsc{G},W\right) \!+\!\mu_{W} \mathrm{tr}\left(W^{T}\!W\right) \label{eq:QCQP}\\
\mathrm{s.t.}\quad &  \mathbb{P}\left(\mathcal{C}_{\mathcal{B}}\geq \zeta \vert x_k\right)\geq p_k, \quad \forall x_k\in \mathcal{D}, \label{eq:p_cb} \\
 &\mathbb{P}\left(\mathcal{C}_L\leq \delta \vert x_k\right)\geq p_k,\quad \forall x_k\in \mathcal{D}, \label{eq:p_cl}  
\end{align}
\end{subequations}and $\forall k\in \mathbb{Z}^+$, where $-\ln p\left(\textsc{X}\vert \textsc{G},W\right)\!=\!\frac{1}{2\sigma^2}\sum_{k=0}^{T_m-1}\Vert x_{k+1}\!-\!W^Tg(s_k)\Vert^2$, $\mu_{W}\in\mathbb{R}^{+}$ is the regularization parameter, and $W^*$ denotes the optimal solution for $W$, which is used in (\ref{eq:compact_NN}) to represent the learned system. 

\begin{prop}
Given the system model in (\ref{eq:compact_NN}) and Assumption \ref{assump_4}, the ELM parameter learning problem with probabilistic safety and stability constraints given in (\ref{eq:QCQP})-(\ref{eq:p_cl}) can be formulated as a QCQP with the objective function in (\ref{eq:QCQP}) and the following constraints:
\begin{subequations}
\begin{align}
  & \Vert \mathcal{A}^{\frac{1}{2}}\left(W^Tg(s_k)-\bar{x}\right) \Vert^2 \leq \Gamma_{\mathcal{C}_{\mathcal{B}}},\label{eq:qcqp_cb} \\
 & \Vert \mathcal{P}^{\frac{1}{2}}\left(W^Tg(s_k)-x^*\right) \Vert^2\leq \Gamma_{\mathcal{C}_L} \label{eq:qcqp_cl} , 
\end{align}
\end{subequations}
where
\begin{align*}
    \mathcal{A}\triangleq&A+4\sigma^2c(p_k)A^2\!, \quad
    \mathcal{P}\triangleq P+4\sigma^2c(p_k)P^2 \\
    \Gamma_{\mathcal{C}_{\mathcal{B}}}\!\triangleq&\ 1\!-\!\zeta\!-\!\sigma^2\mathrm{tr}(A)\!\!-\!(1\!-\!\gamma)h(x_k)\!-\!2\sigma^4\!c(p_k)\mathrm{tr}(A^2)\\
    \Gamma_{\mathcal{C}_L}\!\triangleq &\ \delta - \sigma^2\mathrm{tr}(P)\!\!+\!(1\!-\!\rho)V(x_k)\!-\!2\sigma^4\!c(p_k)\mathrm{tr}(P^2) \\
\end{align*}
\end{prop}
\begin{pf}
Using Assumption \ref{assump_4}, the RHS of (\ref{eq:Prob-DTZBF}) can be lower bounded, and the constraint can be rewritten as
\begin{align}
    \zeta - \mathbb{E}\left[\mathcal{C}_{\mathcal{B}}\right] \leq -c(p_k) V\!ar\left[\mathcal{C}_{\mathcal{B}}\right]. \label{eq:upperbound_CB}
\end{align}
Substituting $\mathbb{E}\left[\mathcal{C}_{\mathcal{B}}\right]$ and $V\!ar\left[\mathcal{C}_{\mathcal{B}}\right]$ from (\ref{eq:E_DZBF}) and (\ref{eq:Var_DZBF}) into (\ref{eq:upperbound_CB}) we have
\begin{align}
&\left\Vert A^{\frac{1}{2}}\left(W^{T}g(s_{k})\!-\!\bar{x}\right)\right\Vert^{2}\!\!\!+\!4c(p_{k})\!\left\Vert \sigma A\left(W^{T}g(s_{k})\!-\!\bar{x}\right)\right\Vert ^{2}\!\leq \nonumber \\
&1\!-\!\zeta\!-\!\sigma^{2}\mathrm{tr}(A)-(1-\gamma)h(x_{k})\!-2\sigma^{4}c(p_{k})\mathrm{tr}(A^{2}), \label{eq:prop_1}
\end{align}
where the RHS of (\ref{eq:prop_1}) can be substituted by $\Gamma_{\mathcal{C}_{\mathcal{B}}}$. Expanding and rearranging (\ref{eq:prop_1}) we get 
\begin{align}
\left(W^{T}\!g(s_{k})\!-\!\bar{x}\right)^{T}\!\!\left(A\!+\!4\sigma^{2}c(p_{k})A^{2}\right)\left(W^{T}g(s_{k})\!-\!\bar{x}\right) & \leq\Gamma_{\mathcal{C}_{\mathcal{B}}}, \label{eq:prop_2}
\end{align}
which after substituting $\mathcal{A}$ for the middle term in (\ref{eq:prop_2}) the result in (\ref{eq:qcqp_cb}) is obtained. Similarly, using Assumption \ref{assump_4} to upper bound the RHS of (\ref{eq:Prob-DTLF}) we have
\begin{align}
    \delta -\mathbb{E}\left[\mathcal{C}_L\right]\geq c(p_k)V\!ar\left[\mathcal{C}_L\right], \label{eq:prop_3}
\end{align}
and when $\mathbb{E}\left[\mathcal{C}_L\right]$ and $V\!ar\left[\mathcal{C}_L\right]$ from (\ref{eq:E_DLF}) and (\ref{eq:Var_DLF}) are substituted into (\ref{eq:prop_3}) we get 
\begin{align}
    &\left\Vert P^{\frac{1}{2}}\left(W^{T}\!g(s_{k})\!-\!x^{*}\right)\right\Vert^{2}\!\!+\!4c(p_{k})\left\Vert \sigma P\!\left(W^{T}g(s_{k})\!-\!x^{*}\right)\right\Vert^{2}\!\! \leq \nonumber\\
& \delta-\sigma^{2}\mathrm{tr}(P)+(1-\rho)V(x_{k})-2\sigma^{4}c(p_{k})\mathrm{tr}(P^{2}), \label{eq:prop_4}
\end{align}
where the RHS of (\ref{eq:prop_4}) can be substituted by $\Gamma_{\mathcal{C}_L}$. Expanding and rearranging (\ref{eq:prop_4}) we get 
\begin{align}
    \left(W^{T}\!g(s_{k})\!-\!x^{*}\!\right)^{T}\!\!\!\left(P\!+\!4\sigma^{2}c(p_{k})P^{2}\right)\left(W^{T}\!g(s_{k})\!-\!x^{*}\right)\!\leq\!\Gamma_{\mathcal{C}_{L}}, \label{eq:prop_5}
\end{align}
which after substituting $\mathcal{P}$ for the middle term in (\ref{eq:prop_5}) the result in (\ref{eq:qcqp_cl}) is obtained. 
\end{pf}

The constraints (\ref{eq:qcqp_cb}) and (\ref{eq:qcqp_cl}) for the optimization problem must be satisfied for all $x_k \in \mathcal{D}$, which leads to a semi-infinite program. In order to circumvent the challenge of having an uncountably infinite set, Theorem \ref{thm:1} is formulated that proves only enforcing a modified constraints for a finite number of sampled points is sufficient for obtaining a solution for the optimization problem in (\ref{eq:QCQP}). 

Let $\mathcal{D}_{\tau}\subset\mathcal{D}$ be a discretization of the state space $\mathcal{D}$ with the closest point in $\mathcal{D}_{\tau}$ to $x_k\in\mathcal{D}$ denoted by $\Vert x_k-[x_k]_{\tau}\Vert\leq\frac{\tau}{2}$, where $\tau$ is the discretization resolution and $\left[\cdot \right]_{\tau}$ denotes the function is evaluated at a point of the discretized space $\mathcal{D}_{\tau}$. Thus, sampling $x_k$ from $\mathcal{D}_{\tau}$, we have $\Vert s_k - \left[s_k\right]_{\tau} \Vert \leq \frac{\tau}{\sqrt{2}}$. 

\begin{thm}\label{thm:1}
If Assumptions \ref{assump_2}-\ref{assump_4} hold, then the optimization problem in (\ref{eq:QCQP})-(\ref{eq:p_cl}) can be solved with the objective in (\ref{eq:QCQP}) and the following modified conditions
\begin{align}
    \Vert \mathcal{A}^{\frac{1}{2}}\!\left(W^T\!\left[g(s_k)\right]_{\tau}\!\!-\!\bar{x}\right)\! \Vert^2 \!\!-\!\! \left[\Gamma_{\mathcal{C}_{\mathcal{B}}}\right]_{\tau} & \!\leq \!-\tau\!\left(\chi_{\mathcal{C}_{\mathcal{B}}}\!+\!\Upsilon_{\mathcal{C}_{\mathcal{B}}}\!\right)\!, \label{eq:qcqp_cb_modified}\\
    \Vert \mathcal{P}^{\frac{1}{2}}\!\left(W^T\!\left[g(s_k)\right]_{\tau}\!\!-\!x^*\right)\! \Vert^2 \!-\! \left[\Gamma_{\mathcal{C}_L}\right]_{\tau} &\! \leq \!-\tau\!\left(\chi_{\mathcal{C}_L}\!+\!\Upsilon_{\mathcal{C}_L}\!\right)\!, \label{eq:qcqp_cl_modified}
\end{align}
$\forall \left[x\right]_{\tau}\in \mathcal{D}_{\tau}$, where 
\begin{subequations}\label{eq:qcqp_bounds}
\begin{align}
    \chi_{\mathcal{C}_{\mathcal{B}}}&\triangleq \left(\lambda_{\mathrm{max}}\{\mathcal{M}\}\sqrt{n_h+1}+\bar{W}\lambda_{\mathrm{max}}\{\mathcal{A}\}\Vert \bar{x}\Vert \right) \bar{g}, \\
    \Upsilon_{\mathcal{C}_{\mathcal{B}}}&\triangleq \left(1-\gamma\right)\lambda_{\mathrm{max}}\{A\}\left(\Vert x_{k}\Vert +\Vert\bar{x} \Vert\right),\\
    \chi_{\mathcal{C}_L}&\triangleq \left(\lambda_{\mathrm{max}}\{\mathcal{H}\}\sqrt{n_h+1}+\bar{W}\lambda_{\mathrm{max}}\{\mathcal{P}\}\Vert x^*\Vert\right)\bar{g}, \\
    \Upsilon_{\mathcal{C}_L}&\triangleq \left(1-\rho\right)\lambda_{\mathrm{max}}\{P\}\left(\Vert x_{k}\Vert +\Vert x^* \Vert\right),\\
    \mathcal{M} &\triangleq W\mathcal{A}W^T, \ \mathcal{H}\triangleq W\mathcal{P}W^T, \ \bar{g} \triangleq \frac{\bar{a}_p\bar{U}\sqrt{n_h}}{2\sqrt{2}} , 
\end{align}
\end{subequations}
$\bar{a}_p = \Vert \mathrm{diag}(a_p) \Vert_F$, and $\lambda_{\mathrm{max}}\{\cdot\}$ denotes the maximum eigenvalue of the argument, then the safety and stability conditions in (\ref{eq:qcqp_cb}) and (\ref{eq:qcqp_cl}) are satisfied $\forall x_k \in \mathcal{D}$. 
\end{thm}
\begin{pf}
The modified safety constraint in (\ref{eq:qcqp_cb_modified}) can be written as 
\begin{align}
    \Vert \mathcal{A}^{\frac{1}{2}}\left(W^T\left[g\right]_{\tau}\!\!-\!\bar{x}\right)\! \Vert^2 \!-\! \left[\Gamma_{\mathcal{C}_{\mathcal{B}}}\right]_{\tau}  \!\!+\tau\left(\chi_{\mathcal{C}_{\mathcal{B}}}\!+\!\Upsilon_{\mathcal{C}_{\mathcal{B}}}\right) &\! \leq 0, \label{eq:thm_1}
\end{align}
$\forall x_k \in \mathcal{D}$. If $\Vert \mathcal{A}^{\frac{1}{2}}\left(W^Tg-\bar{x}\right) \Vert^2 \!-\! \Gamma_{\mathcal{C}_{\mathcal{B}}}$ is a lower bound on the left hand side of (\ref{eq:thm_1}) then the safety constraint in (\ref{eq:qcqp_cb}) is satisfied. Consider 
\begin{align*}
    \mathcal{O}_{\mathcal{B}}\!\!=\!\!\Vert\mathcal{A}^{\frac{1}{2}}\!\!\left(W^T\!g\!-\!\bar{x}\right)\!\Vert^2\!\!-\! \Vert \mathcal{A}^{\frac{1}{2}}\!\!\left(W^T\!\left[g\right]_{\tau}\!\!-\!\bar{x}\right)\! \Vert^2\!-\!\!\left(\Gamma_{\mathcal{C}_{\mathcal{B}}}\!\!-\!\left[\Gamma_{\mathcal{C}_{\mathcal{B}}}\right]_{\tau}\right)\!,
\end{align*}
which by using the triangle inequality, $\mathcal{O}_{\mathcal{B}}$ can be upper bounded as
\begin{align*}
    \mathcal{O}_{\mathcal{B}}\!\!\leq\!\left\vert\Vert\mathcal{A}^{\frac{1}{2}}\!\!\left(W^T\!g\!-\!\bar{x}\right)\!\Vert^2 \!\!-\!\! \Vert \mathcal{A}^{\frac{1}{2}}\!\!\left(W^T\!\left[g\right]_{\tau}\!\!-\!\bar{x}\right)\! \Vert^2\right\vert\!\!+\!\!\big\vert\Gamma_{\mathcal{C}_{\mathcal{B}}}\!\!-\!\!\left[\Gamma_{\mathcal{C}_{\mathcal{B}}}\right]_{\tau}\!\!\big\vert.
\end{align*}
After expanding each term and using Remark \ref{rem:1}, we have
\begin{align*}
&\left\vert \Vert\mathcal{A}^{\frac{1}{2}}\!\!\left(W^{T}\!g\!-\!\bar{x}\right)\!\Vert^{2}\!-\!\Vert\mathcal{A}^{\frac{1}{2}}\!\!\left(W^{T}\!\left[g\right]_{\tau}\!\!-\!\bar{x}\right)\!\Vert^{2}\right\vert \!+\!\big\vert\Gamma_{\mathcal{C}_{\mathcal{B}}}\!-\!\left[\Gamma_{\mathcal{C}_{\mathcal{B}}}\right]_{\tau}\!\big\vert\!\leq\\
&\left(\lambda_{\mathrm{max}}\left\{\mathcal{M}\right\}\sqrt{n_h+1}+\bar{W}\lambda_{\mathrm{max}}\!\left\{ \mathcal{A}\right\} \left\Vert \bar{x}\right\Vert \right)\frac{\bar{a}_p\bar{U}\sqrt{n_h}}{2\sqrt{2}}\tau+\\
&\left(1\!-\!\gamma\right)\lambda_{\mathrm{max}}\left\{A\right\}\left(\left\Vert x_{k}\right\Vert + \left\Vert \bar{x}\right\Vert \right)\tau,
\end{align*} which yields 
\begin{align}
&\Vert\mathcal{A}^{\frac{1}{2}}\!\left(W^{T}\!g\!-\!\bar{x}\right)\!\Vert^{2}\!\!-\Gamma_{\mathcal{C}_{\mathcal{B}}}\leq\Vert\mathcal{A}^{\frac{1}{2}}\!\left(W^{T}\!\left[g\right]_{\tau}\!-\!\bar{x}\right)\!\Vert^{2}\!-\!\left[\Gamma_{\mathcal{C}_{\mathcal{B}}}\right]_{\tau}\!+\!\nonumber \\
&\left(\lambda_{\mathrm{max}}\left\{\mathcal{M}\right\}\sqrt{n_h+1} +\bar{W}\lambda_{\mathrm{max}}\left\{\mathcal{A}\right\} \left\Vert \bar{x}\right\Vert \right)\bar{g}\tau + \left(1\!-\!\gamma\right)\!\times \nonumber \\
&\lambda_{\mathrm{max}}\!\left\{A\right\}\left(\left\Vert x_{k}\right\Vert\!+\!\left\Vert \bar{x}\right\Vert \right)\tau, \ \forall x_k\in \!\mathcal{D}\ \mathrm{and}\  \forall [x_k]_{\tau} \!\in \mathcal{D}_{\tau}. \label{eq:thm_2}
\end{align}
From (\ref{eq:thm_1}), (\ref{eq:thm_2}), and (\ref{eq:qcqp_bounds}) it is concluded that \newline $\Vert\mathcal{A}^{\frac{1}{2}}\left(W^{T}g-\bar{x}\right)\!\Vert^{2}-\Gamma_{\mathcal{C}_{\mathcal{B}}}\leq 0$, $\forall x_k\in \mathcal{D}$.

In a similar fashion, the modified stability constraint in (\ref{eq:qcqp_cl_modified}) can be written as 
\begin{align}
    \Vert \mathcal{P}^{\frac{1}{2}}\left(W^T\left[g\right]_{\tau}\!\!-\!x^*\right)\!\Vert^2 \!-\! \left[\Gamma_{\mathcal{C}_L}\right]_{\tau}  \!\!+\!\tau\left(\chi_{\mathcal{C}_L}\!+\!\Upsilon_{\mathcal{C}_L}\right) &\! \leq 0, \label{eq:thm_cl1}
\end{align}
$\forall x_k \in \mathcal{D}$. If $\Vert \mathcal{P}^{\frac{1}{2}}\left(W^Tg-x^*\right) \Vert^2 \!-\! \Gamma_{\mathcal{C}_L}$ is a lower bound on the left hand side of (\ref{eq:thm_cl1}) then the safety constraint in (\ref{eq:qcqp_cl}) is satisfied. Consider 
\begin{align*}
    \mathcal{O}_L\!\!=\!\Vert\mathcal{P}^{\frac{1}{2}}\!\!\left(W^T\!\!g\!-\!x^*\!\right)\!\Vert^2 \!\!-\! \Vert \mathcal{P}^{\frac{1}{2}}\!\!\left(W^T\!\!\left[g\right]_{\tau}\!\!\!-\!x^*\!\right)\! \Vert^2\!-\!\!\left(\Gamma_{\mathcal{C}_L}\!\!\!-\!\left[\Gamma_{\mathcal{C}_L}\right]_{\tau}\!\right)\!,
\end{align*}
which by using the triangle inequality, $\mathcal{O}_L$ can be upper bounded as
\begin{align*}
    \mathcal{O}_L\!\!\leq\!\left\vert\Vert\mathcal{P}^{\frac{1}{2}}\!\!\left(W^T\!g\!-\!x^*\!\right)\!\!\Vert^2 \!\!-\!\! \Vert \mathcal{P}^{\frac{1}{2}}\!\!\left(W^T\!\left[g\right]_{\tau}\!\!-\!x^*\!\right)\!\! \Vert^2\right\vert\!\!+\!\!\big\vert\Gamma_{\mathcal{C}_L}\!\!\!-\!\!\left[\Gamma_{\mathcal{C}_L}\right]_{\tau}\!\!\big\vert.
\end{align*}
After expanding each term and using Remark \ref{rem:1}, we have
\begin{align*}
&\left\vert \Vert\mathcal{P}^{\frac{1}{2}}\!\!\left(W^{T}\!g\!-\!x^*\!\right)\!\Vert^{2}\!-\!\Vert\mathcal{P}^{\frac{1}{2}}\!\!\left(W^{T}\!\left[g\right]_{\tau}\!\!-\!x^*\!\right)\!\Vert^{2}\right\vert \!+\!\big\vert\Gamma_{\mathcal{C}_L}\!\!\!-\!\left[\Gamma_{\mathcal{C}_L}\right]_{\tau}\!\big\vert\!\leq\\
&\left(\lambda_{\mathrm{max}}\left\{\mathcal{M}\right\}\sqrt{n_h+1}+\bar{W}\lambda_{\mathrm{max}}\!\left\{ \mathcal{P}\right\} \left\Vert x^*\right\Vert \right)\frac{\bar{a}_p\bar{U}\sqrt{n_h}}{2\sqrt{2}}\tau+\\
&\left(1\!-\!\gamma\right)\lambda_{\mathrm{max}}\left\{P\right\}\left(\left\Vert x_{k}\right\Vert + \left\Vert x^*\right\Vert \right)\tau,
\end{align*}which yields 
\begin{align}
&\Vert\mathcal{P}^{\frac{1}{2}}\!\left(W^{T}\!g\!-\!x^*\right)\!\Vert^{2}\!\!-\Gamma_{\mathcal{C}_L}\leq\Vert\mathcal{P}^{\frac{1}{2}}\!\left(W^{T}\!\left[g\right]_{\tau}\!-\!x^*\right)\!\Vert^{2}\!-\!\left[\Gamma_{\mathcal{C}_L}\right]_{\tau}\!\!+\nonumber \\
&\left(\lambda_{\mathrm{max}}\left\{\mathcal{H}\right\}\sqrt{n_h+1} +\bar{W}\lambda_{\mathrm{max}}\left\{\mathcal{P}\right\} \left\Vert x^*\right\Vert \right)\bar{g}\tau + \left(1\!-\!\rho\right)\!\times \nonumber \\
&\lambda_{\mathrm{max}}\!\left\{P\right\}\left(\left\Vert x_{k}\right\Vert\!+\!\left\Vert x^*\right\Vert\right)\!\tau\!, \ \forall x_k\in \!\mathcal{D}\ \mathrm{and}\  \forall [x_k]_{\tau} \!\in \mathcal{D}_{\tau}. \label{eq:thm_cl2}
\end{align}
From (\ref{eq:thm_cl1}), (\ref{eq:thm_cl2}), and (\ref{eq:qcqp_bounds}), it is concluded that $\Vert\mathcal{P}^{\frac{1}{2}}\!\!\left(W^{T}\!g\!-\!x^*\right)\!\Vert^{2}\!\!-\Gamma_{\mathcal{C}_L}\leq 0$, $\forall x_k\in \mathcal{D}$.
\end{pf}
\begin{rem}\label{rem:4}\normalfont
Theorem \ref{thm:1} proves that enforcing a modified stability and Lyapunov constraints for a finite number of sampled points is sufficient to obtain an optimal solution of the QCQP defined in (\ref{eq:QCQP}).
\end{rem}
\begin{figure}
   \centering
    \includegraphics[width=1\columnwidth]{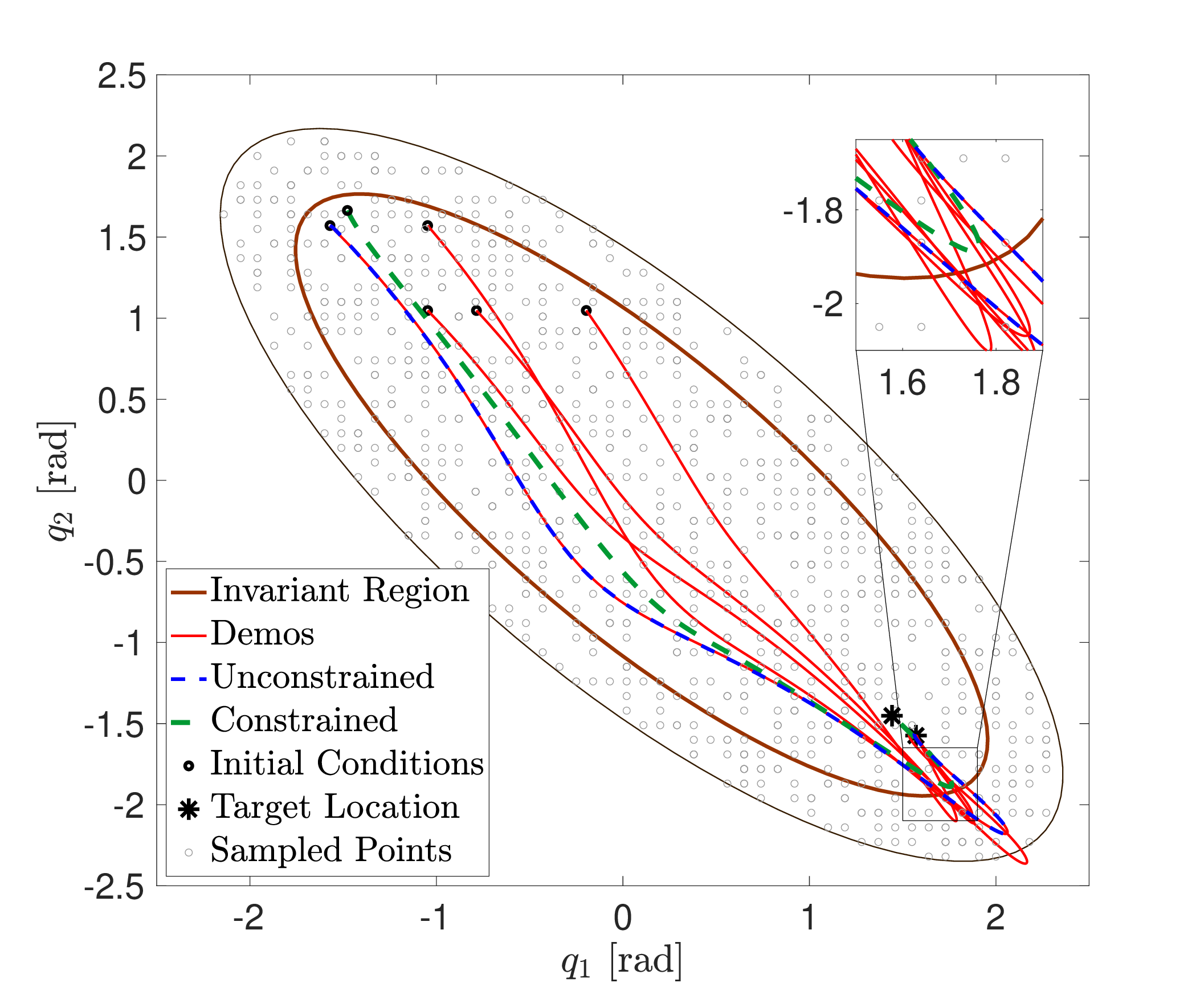}
    \caption{Illustration of the ELM model learning using probabilistic safety and stability constraints on joint positions data of a $2$ DoF planar robot manipulator.\label{fig:ellipse_q}}
\end{figure}

\section{Numerical Evaluations}\label{sec:simulation}
In this section two numerical examples of the proposed method are presented. The optimization problem is solved using the CVX package in MATLAB $2020b$. Random initialization process in the ELM algorithm may lead to having saturated or constant neurons, which are not desired while learning the model \cite{neumann2013optimizing}. To circumvent this problem, 
a batch intrinsic plasticity (BIP) learning rule is used to estimate the slopes and biases of the ELM \cite{neumann2013optimizing}. The active sampling strategy introduced in \cite{salehi2021dynamical} is also used to select informative points for the constraint computations of the optimization problem in (\ref{eq:QCQP}), (\ref{eq:qcqp_cb_modified}), (\ref{eq:qcqp_cl_modified}). For both examples, the risk tolerance of $p_k=0.9$, the variance of $\epsilon$ of $\sigma=0.02$, and regularization parameter $\mu_W = 0.01$ are chosen. The subsequent results are generated using an ELM network with hidden neurons $n_h= 25$. For both examples $100$ Monte Carlo simulation runs are conducted to test the model on bound violations during the trajectory reproduction from various initial conditions.  

In the first example a two-link robot planar manipulator with lengths $L_1=1\ [\mathrm{m}]$ and $L_2=1 \ [\mathrm{m}]$ and masses $m_1 = 1\ [\mathrm{kg}]$ and $m_2=1\ [\mathrm{kg}]$ is considered. Consider the Euler-Lagrange (EL) dynamics $M(q)\ddot{q}+C(q,\dot{q})\dot{q}+G(q)=u$, where $M(q)\in\mathbb{R}^{2\times 2}$ denotes the
inertia matrix, $C(q,\dot{q})\in\mathbb{R}^{2\times 2}$ denotes the centripetal-Coriolis matrix, $G(q)\in\mathbb{R}^{2}$
denotes the gravity vector, $u \in\mathbb{\mathbb{R}}^{2}$
represents the control input vector, and $q, \dot{q} \in\mathbb{R}^{2}$
denote the joint angles and angular velocity, respectively. 
A nonlinear black-box model of the EL-dynamics with only position states is considered, and it is given by $q_{k+1} = f(q_k,u_k)$. 

\begin{figure}
   \centering
    \includegraphics[width=1\columnwidth]{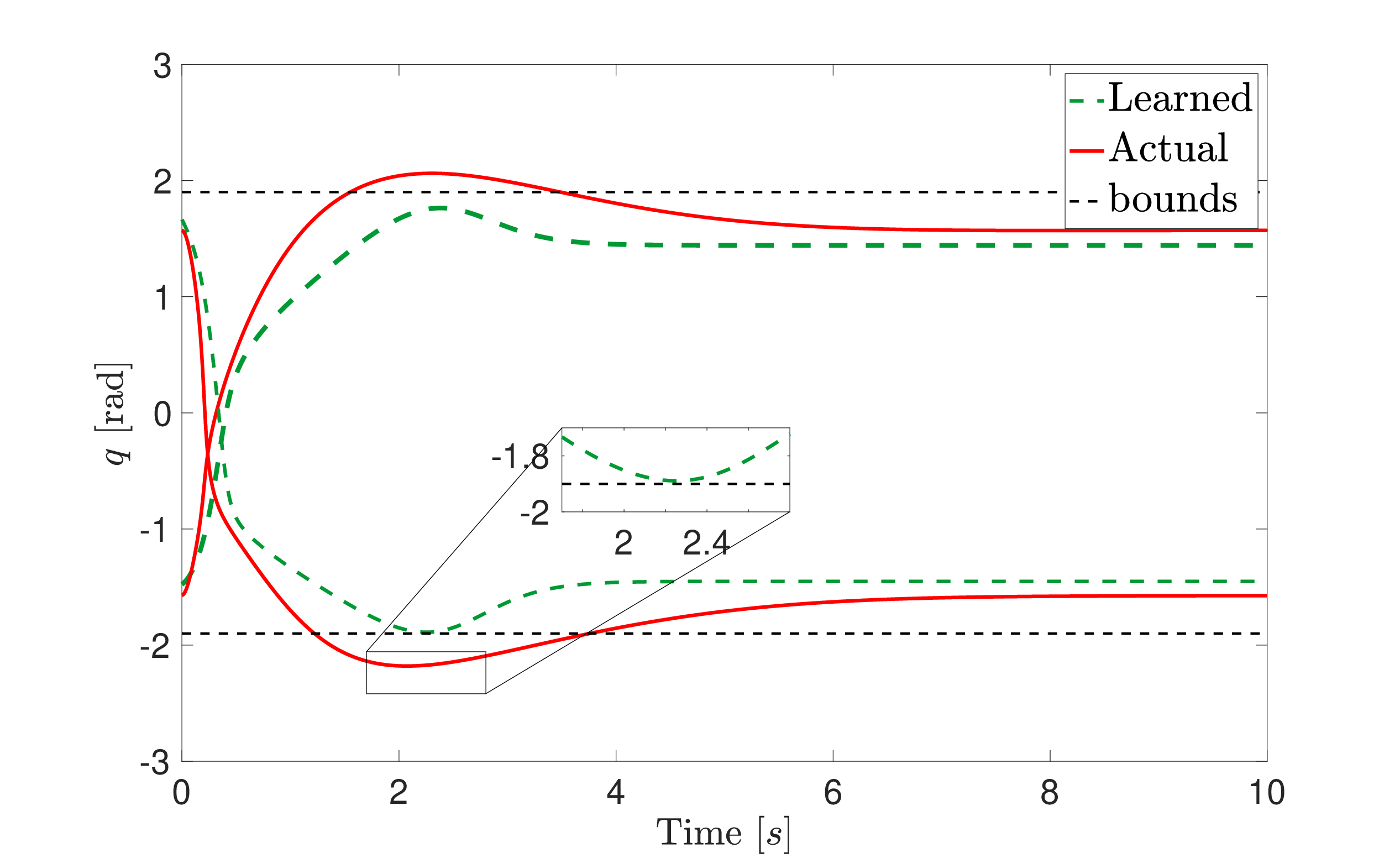}
    \caption{Evolution of the joint angles for the planar robot simulation using the learned ELM parameters.\label{fig:q}}
\end{figure}

To generate the state trajectories data $q_k$ for training the ELM model, a PID set-point controller is designed to regulate the joint positions to a desired value $q=[\frac{\pi}{2},-\frac{\pi}{2}]$. Through empirical investigations, a set of five trajectories with randomly selected initial conditions are collected and their PID gains are tuned individually. 
The joint position satisfies the following constraints: $|q(1)|\leq 1.90$ and $|q(2)|\leq 1.90$, where $q(1)$ and $q(2)$ denote the joint's first and second dimension. 
An invariant region in the form of an ellipse is selected, i.e., $h(q_k)=1-\left(q_k-\bar{q})^TA(q_k-\bar{q}\right)$, where $\bar{q}\in\mathbb{R}^2$ denotes the hyperellipsoid's center of gravity, 
\begin{align}
A & \!=\!\begin{bmatrix}\frac{\cos^{2}\alpha}{\iota_{1}^{2}}\!+\!\frac{\sin^{2}\alpha}{\iota_{2}^{2}} & \cos\alpha\sin\alpha\left(\frac{1}{\iota_{1}^{2}}\!-\!\frac{1}{\iota_{2}^{2}}\right)\\
\cos\alpha\sin\alpha\left(\frac{1}{\iota_{1}^{2}}\!-\!\frac{1}{\iota_{2}^{2}}\right) & \frac{\sin^{2}\alpha}{\iota_{1}^{2}}\!+\!\frac{\cos^{2}\alpha}{\iota_{2}^{2}}
\end{bmatrix},
\end{align}  
$\iota_{1}$ and $\iota_{2}$ are the major and minor axes of the ellipse, respectively and $\alpha$ is the orientation of the ellipse. Parameters of the ellipse are chosen such that the ellipse encloses the demonstrations data and meets the constraints on the joint positions. Other parameters $\gamma = 0.9$, $\rho = 0.01$, $\zeta = 0.01$, and $\delta = 0.01$ are selected empirically. In Fig. \ref{fig:ellipse_q} the results of the ELM parameter learning are shown when the system parameters are subject to probabilistic safety and stability constraints given in (\ref{eq:qcqp_cb_modified}) and (\ref{eq:qcqp_cl_modified}). Using the DT-ZCBF implies asymptotic stability of set $\mathcal{S}$, shown as an invariant region in Fig. \ref{fig:ellipse_q}. Hence, if any bounded disturbances push the state outside the invariant region, the set $\mathcal{S}$ is asymptotically reached. For this reason, when training the system model, the points are sampled from a more extensive set than the selected ellipse \cite{xu2015robustness}. The joint position trajectory is produced at a different initial condition selected at random from a uniform distribution between the initial condition used during training and a point that is $0.5$ millimeters distant apart. It can be seen from Fig. \ref{fig:ellipse_q}, the reproduced trajectory remains inside the desired invariant region for all time and converges to the desired set-point. Same results are observed for all the Monte Carlo runs. Fig. \ref{fig:q} shows the evolution of the joint angles using the learned ELM parameters plotted as function of time. The reproduced joint position trajectories never cross the set boundaries in either direction and asymptotically converge to a point very close to the desired position. 
\begin{figure}
   \centering
    \includegraphics[width=1\columnwidth]{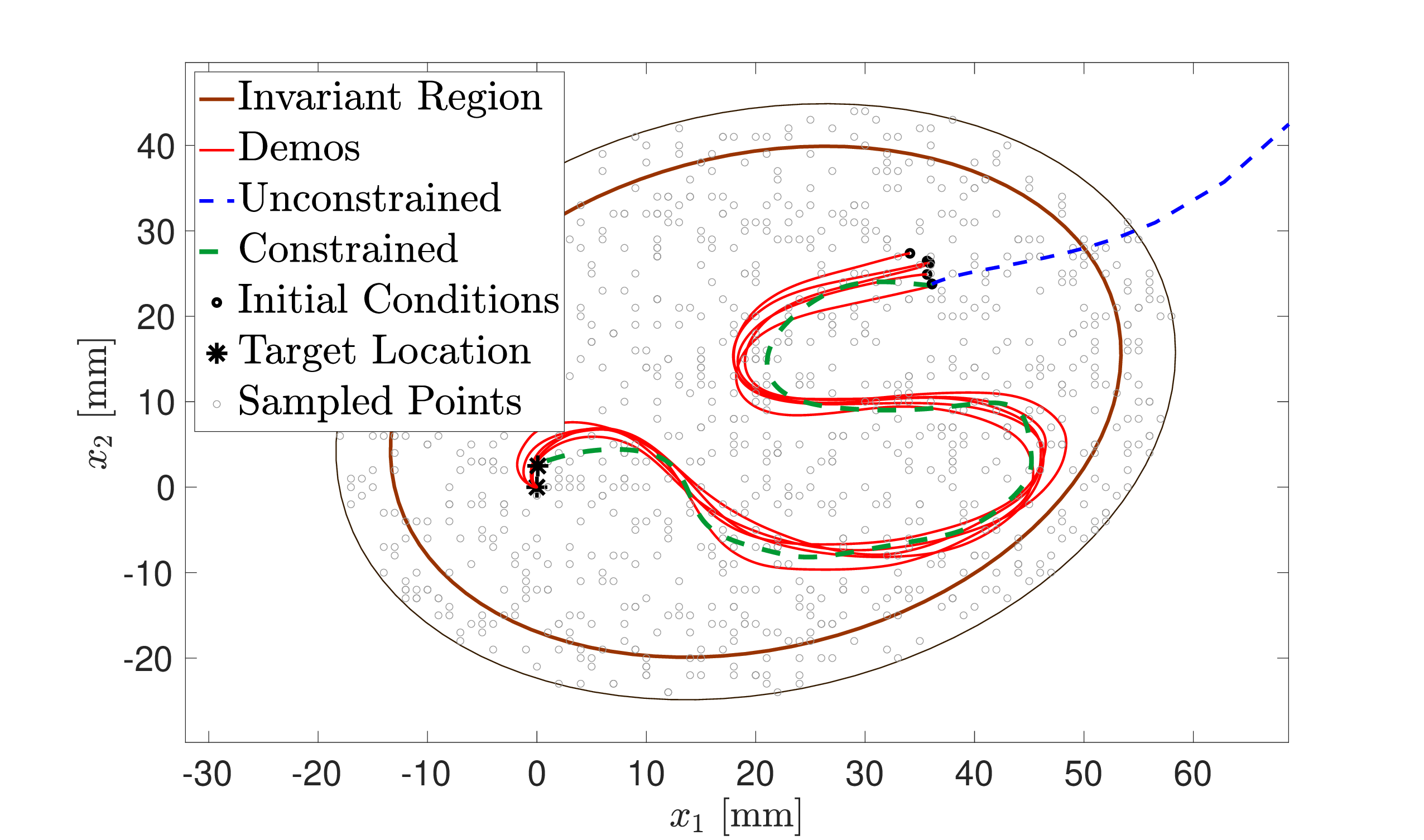}
    \caption{Model \textit{Snake}, learned with probabilistic safety and stability constraints for ellipse sets.\label{fig:LASA}}
\end{figure}

For the second example, the proposed method is tested on the \textit{Snake} shape from the LASA human handwriting dataset \cite{lemme2015open}. An ellipse is chosen for the invariant region that encloses the Snake shape's $7$ demonstrations. The state trajectory data are translated by a constant before training so that the error between the current state and the target location can be used as an input to the network instead of the control signal. Once the model is learned the state trajectories are translated back to zero. The optimization design parameters for this simulation are as follows: $\gamma = 0.9$ and $\rho = 0.3$, $\zeta = 0.1$, $\delta = 1$, and they are selected empirically. Fig. \ref{fig:LASA} shows the trajectories generated by the learned model remain inside a safe set, and converge to the system equilibrium. Since the constraints are derived for one-step-ahead prediction, no barrier violation occurs during the Monte Carlo runs if the initial condition is sufficiently close to the initial condition of the demonstration trajectories. The average CPU time for training the ELM network for both simulations is computed over five independent runs with a $1000$ number of sampling points is $2.32$ seconds. 

\section{Conclusion}

A nonlinear system identification method using the ELM with probabilistic DT-ZCBF and DT-CLF constraints is presented. The ELM reconstruction error is modeled as a normally distributed Gaussian noise, which induces a distribution over the safety and stability constraints. Therefore, the probabilistic form of the ZBF and Lyapunov constraints are derived such that safety and stability are guaranteed within user-defined risk tolerance. A theorem is developed to make the QCQP implementable, which proves that with a modification to the constraints, the quadratic program with an infinite number of constraints can be relaxed to a finite number of constraints. The applicability of the proposed model learning method is demonstrated using two practical robotics engineering examples.



\bibliographystyle{unsrt}

\bibliography{IFAC}           



\end{document}